\documentclass{aa}
\usepackage{graphicx,natbib}
\usepackage[latin1]{inputenc}
\usepackage[T1]{fontenc}
\usepackage{txfonts}
\usepackage{changebar}
\usepackage{bm}
\usepackage{units}

\def\mmol{M_{\rm Mol}}
\def\tetad{\theta_{\rm D}}

\def\mearth{M_{\oplus}}
\def\mocean{M_{\rm Ocean}}
\def\rearth{R_{\oplus}}

\def\f1{f_{\rm I}}

\def\beq{\begin{equation}}
\def\eeq{\end{equation}}

\def\mearth{M_\oplus}

\def\mearth{M_\oplus}

\def\simgr{\,\hbox{\hbox{$ > $}\kern -0.8em \lower 1.0ex\hbox{$\sim$}}\,}
\def\simle{\,\hbox{\hbox{$ < $}\kern -0.8em \lower 1.0ex\hbox{$\sim$}}\,}
\def\beq{\begin{equation}}
\def\eeq{\end{equation}}

\def\simgr{\,\hbox{\hbox{$ > $}\kern -0.8em \lower 1.0ex\hbox{$\sim$}}\,}
\def\simle{\,\hbox{\hbox{$ < $}\kern -0.8em \lower 1.0ex\hbox{$\sim$}}\,}
\def\beq{\begin{equation}}
\def\eeq{\end{equation}}

\def\apj{ApJ}                 
\def\apjl{ApJ}                
\def\apjs{ApJS}               
\def\ao{Appl.Optics}          
\def\aap{A\&A}                

\def\({\left(}
\def\){\right)}
\def\<{\left<}
\def\>{\right>}

\def\({\left(} 
\def\){\right)} 
\def\<{\left<} 
\def\>{\right>} 

\def\bc{\begin{changebar}}
\def\bce{\begin{center}}
\def\beq{\begin{equation}} 

\def\bi{\begin{itemize}}

\def\btab{\begin{tabular}{p{1.7cm}p{12cm}p{1.5cm}}}
\def\bt2{\begin{tabular}{p{1 cm}p{4.5cm}p{10cm}}}

\def\ec{\end{changebar}}
\def\ece{\end{center}}
\def\eeq{\end{equation}} 
\def\ei{\end{itemize}}

\def\etab{\end{tabular}\\}

\def\mH2{m_\mathrm{H_2}}

\def\dmax0{\rho_\mathrm{max}}
\def\dmaxS0{\Sigma_\mathrm{max}}

\def\rH2{r_\mathrm{H_2}}

\def\r0max{r_\mathrm{0max}}

\def\s0{\sigma_\mathrm{0}}

\def\xp0{x_{\rm{M0}}}

\def\z0max{z_\mathrm{0max}}

\def\apj{{\it ApJ}}                 
\def\apjl{ApJ}                
\def\apjs{ApJS}               
\def\ao{Appl.Optics}          
\def\aap{{\it A\&A}}                
\def\jgr{J. Geophys. Res} 

\begin{document}

\title{On the  radius of habitable planets}

\author{Y. Alibert \inst{1,2}}
\offprints{Y. Alibert}
\institute{Physikalisches Institut  \& Center for Space and Habitability, Universitaet Bern, CH-3012 Bern, Switzerland, 
        \email{yann.alibert@space.unibe.ch}
           \and
        Observatoire de Besan\c con, 41 avenue de l'Observatoire, 25000 Besan\c con, France}

\abstract
{The conditions that a planet must fulfill to be habitable are not precisely known. However, it is comparatively easier to define conditions under which a planet
is very likely not habitable. Finding such conditions is important as it can help select, in an ensemble of potentially observable planets, which ones should
 be observed in greater detail for characterization studies.}
{Assuming, as in the Earth, that the presence of a C-cycle is a necessary condition for long-term habitability, we derive, as a function of the planetary mass,
a radius above which a planet is likely not habitable.  We compute the maximum radius a planet can have to fulfill two constraints: surface conditions compatible 
with the existence of liquid water, and no ice layer at the bottom of a putative global ocean. We demonstrate that, above a given radius, these two constraints cannot be met.}
{We compute internal structure models of planets, using a five-layer model (core, inner mantle, outer mantle, ocean, and atmosphere), for different masses and composition of the planets (in particular, the Fe/Si ratio of the planet).}
{Our results show that for planets in the Super-Earth mass range (1-12 $\mearth$), the maximum that a planet, with a composition similar to that of the Earth, can have varies between 1.7 and 2.2 $\rearth$. This radius is reduced when considering planets with higher Fe/Si ratios and taking radiation into account when computing the gas envelope structure.}
{These results can  be used to infer, from radius and mass determinations using high-precision transit observations like those that will soon be performed by the CHaracterizing ExOPlanet Satellite (CHEOPS), which planets are very likely \textit{not} habitable, and therefore which ones should be considered as best targets for further habitability studies. } 

\maketitle

\section{Introduction}
\label{sec:introduction}

Since the discovery of the first extrasolar planet around a main-sequence star (Mayor and Queloz 1995), observational
programs have led to the discovery of lower and lower mass planets (Mayor et al. 2011), some of them located in the so-called habitable zone at
a location where it is believed that liquid water could exist on the planetary surface. The possible presence of liquid
water is in turn believed to be important for the habitability of planets, as it is required for life (as we know it) to 
exist and develop. 

The habitability of planets is - however - an ill-defined concept, and there is nowadays no clear definition of it. Indeed, the
presence of liquid water only, while it could be a necessary condition, is probably not enough for habitability. What is,
on the other hand, quite easy to define, is non-habitability, and one may easily find different conditions under which a planetary
surface is very likely not habitable. In the context of habitability studies, the definition of such non-habitability conditions
is useful, for example as it may allow to select future observational targets when large-scale facilities  allow us
to directly characterize in great detail the most promising extrasolar planets. The goal of this study is to provide a clear
non-habitability criterion, that can be easily derived once the mass and the radius of an extrasolar planet (located for example
in the habitable zone) is known. In addition, we will see that the simultaneous determination of the central star composition, as well as the observation of Rayleigh scattering in the planetary atmosphere, can be used to derive stronger constraints on planetary habitability.

Indeed, it has been recognized for many years that one key element for habitability, at least in the case of the Earth, but
also likely for many planets, is the presence of the Carbon cycle (see e.g., Kasting 2010). Indeed, on Earth, the C-cycle acts as 
a very important temperature stabilizing process and buffers the surface temperature at values close to those allowing liquid water. This is
especially important since the luminosity of a star increases with time and, without any stabilization process, it could be
difficult to maintain liquid water at a surface of any planet during an appreciable amount of time.

The C-cycle has been presented in many papers and books (see e.g., Kasting 2010; Pierrehumbert 2010), and we will not describe it in  detail. The key 
process during this C-cycle is the weathering of silicates, which converts silicates to carbonates, as a result
of the interaction between atmospheric CO$_2$ dissolved in water and silicates of the planetary rocky surface. Carbonates then precipitate at
the bottom of oceans and can be engulfed in subduction zones and transported toward high-temperature regions of planetary
mantle. The cycle is then completed as carbonates are destabilized at high temperature, producing CO$_2$ that can finally be
sent back to the atmosphere by volcanoes. 

The stabilization effect of the C-cycle comes from both the dependence of silicate weathering on temperature\footnote{Silicate weathering is a growing function
of temperature and can be stopped in the case of a planet covered by ice (see Pierrehumbert 2010).} and the strong greenhouse effect of CO$_2$. If the surface temperature
decreases, the removal of CO$_2$ through silicate weathering is suppressed, while CO$_2$ is still released into the atmosphere by volcanoes.
This increases the CO$_2$ concentration in the atmosphere and eventually the temperature. On the contrary, if the surface temperature
increases, then silicate weathering increases, which decreases the CO$_2$ atmospheric concentration and greenhouse effect, thus lowering the surface
temperature. Silicate weathering therefore requires a reaction between CO$_2$ dissolved
in oceans, or directly with CO$_2$ present in the atmosphere, and rocks from the planetary mantle. This, in turn, is only possible if there is a physical interface
between liquid water (or atmosphere) and rocks.

Water-rich planets (or ocean planets) are covered by a global ocean. If the amount of water is large enough (the exact value depending
on the planetary gravity), the pressure at the bottom of the global ocean is so large that a layer of high pressure ice (ice VII) appears\footnote{Note that, contrary to low pressure ice, 
ice VII has a density that is higher than the one of liquid water and stays at the bottom of the ocean.}.
This effectively prevents any contact between silicates and liquid water (and therefore atmospheric CO$_2$ that could be dissolved in the
ocean), and is likely to suppress silicate weathering. As a result, the C-cycle and its stabilization effect cannot exist. The working hypothesis of this study is therefore that a necessary condition for habitability is the presence of C-cycle. As a consequence,
planets with a high enough water content, and therefore a high pressure ice layer at the bottom of a global ocean, are considered  not habitable\footnote{It should be noted that 
recent studies (Abbot et al. 2012) have shown that silicate weathering requires some land to be
present on the planetary surface. As a consequence, according to these authors,  a tiny amount of water (large enough to cover
 the entire surface) is enough to prevent habitability, at least under our working hypothesis.}.

\begin{figure}
  \center
  \includegraphics[height=0.25\textheight]{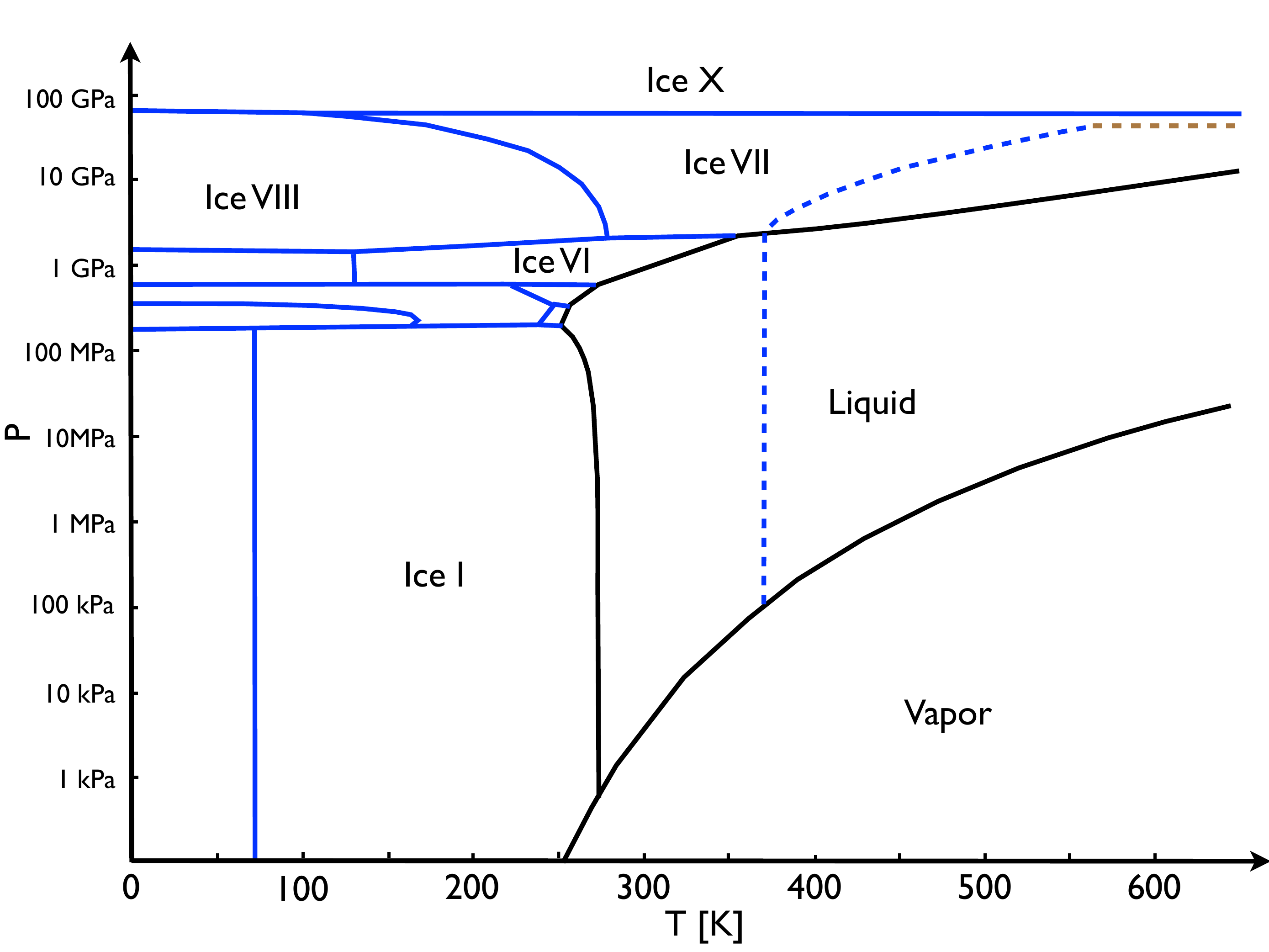}
  \caption{Simplified phase diagram of water, and thermodynamical properties inside a 5 $\mearth$ planet, containing 0.5 $\mearth$ of water, with surface
  temperature equal to 370K and surface pressure at the vapour/liquid transition. The Fe/Si and Mg/Si ratio are equal to the solar values. 
  The temperature and pressure are plotted in brown for the silicate/iron phase, and in blue for the water. As can be seen on the figure, this planet
  presents a large layer of high pressure ice, preventing habitability.}
  \label{phase_diagram}
\end{figure}

The radius of a planet depends mainly on its composition. The effect of the temperature profile has been shown to have a small influence
on it (see e.g., Sotin et al. 2007, Seager et al. 2007, Grasset et al. 2010, Valencia et al. 2010). For a given mass, a large radius implies either the presence of a lot of water, or the presence of a 
(relatively) massive gaseous envelope, or both. In the first case, as we have argued before, habitability is hindered by the presence of a high pressure 
ice layer at the bottom of the global ocean. In the second case, the temperature and pressure at the bottom of the atmosphere are too large to allow for
the presence of liquid water. In both cases, the planet is therefore likely not habitable, and will be considered accordingly in this paper. 
The process is exemplified in Fig. \ref{phase_diagram}, which presents a simplified phase diagram of water, together with a model of a $5 \mearth$ water-rich planet.
This planet presents a thick layer of high pressure ice at the bottom of the ocean, which prevents the C-cycle from operating.

The goal of this study is to compute the maximum radius that a planet can have in order that 1) the pressure at the bottom of the ocean
is low enough to prevent the apparition of high pressure ice and 2) the temperature and pressure at the surface of the planet is
compatible with liquid water. It is important to note that such a calculation does not constrain in any way the past or future habitability
of a given planet. In particular, following the arguments of Abbot et al. (2012), the suppression of silicate weathering on a planet
may lead to a runaway greenhouse effect and massive loss of water. As a consequence, such a planet could become habitable in the future.
However, the same planet would not be habitable today.

The paper is organized as follows: we present in Sect. \ref{model} our internal structure model of planetary interior and planetary atmosphere, and some validation test we conducted in Sect. \ref{1mearth}.
In Sect. \ref{max_all}, we compute the maximum radius of habitable planet under rather extreme hypothesis on the planetary bulk composition (in particular for planets without Fe). 
This unlikely planetary composition is used to provide the overall maximum of planets, whatever the
composition. In Sect. \ref{max_fe}, assuming different Fe/Si ratios (that may reflect the composition of the central star), we compute the corresponding maximum
radius. Finally, Sect. \ref{conclusion} is devoted to our discussion and conclusions.

\section{Model}
\label{model}

We compute the internal structure of planets that consists of five layers: a core, an inner mantle, an outer mantle, a water layer, and a gas envelope.
The presence of a core depends on the assumed Fe/Si ratio in the planet. Indeed, for a given composition, the largest radius is obtained when the maximum of Fe is present in silicates. For each planetary composition,
we therefore compute the structure with the smallest possible core, since we want to derive an upper boundary of the planetary radius. It appears that, for solar Mg/Si and Fe/Si ratios, all the available Fe 
can be present in the inner and outer mantle and no core is required to match the composition. For Fe/Si > 1.13 Fe/Si$ )_\odot$ (for the chemical species we consider in the model), this is not any more possible and planets have an iron core. We emphasize
the fact that assuming undifferentiated planets, as we do here, is probably far from the reality, at least for planets more massive than the Earth. However, we make this assumption as it provides us
with an upper boundary of the possible planetary radii for a given mass and composition.

Our model of the four (three, if no core is required) innermost layers closely follows the model proposed by Sotin et al. (2007) and further improved by Grasset et al. (2010). The core is made of 
Fe (for simplicity, we do not include any inner/outer core dichotomy and do not include the effect of the presence of a volatile like S in the core), the inner mantle is made of perovskite (MgSiO$_3$/FeSiO$_3$) 
and wustite (MgO/FeO), the outer mantle is made of olivine (Mg$_2$SiO$_4$/Fe$_2$SiO$_4$) and entstatite (Mg$_2$Si$_2$O$_6$/Fe$_2$Si$_2$O$_6$). We assume the water mantle is  
made of pure water, and the planetary envelope is made of either H$_2$ or H/He in solar composition, these latter being treated as ideal gases. We also
consider other gaseous compositions  to study their effect on the maximum radius. This choice of the planetary  atmospheric composition is dictated by our goal to derive a maximum planetary envelope.
As we will see later, other choices of atmospheric composition  produce smaller radii.

We solve the standard internal structure equations: 
\begin{equation}
{d r \over d P } = { 1 \over \rho g }
\end{equation}
\begin{equation}
{d m \over d P } = { 4 \pi r^2 \over g }
\end{equation}
and
\begin{equation}
{d T \over d P } =  {T \over P}  \nabla_{\rm ad}
\end{equation}
where $P$ is the pressure, $r$ the radius, $m$ the mass interior to radius $r$, $g$ the gravity, $\rho$ the density given by the equation of state (see below), $T$ the temperature,
and $\nabla_{\rm ad} = {(d \ln T / d \ln P)}_{\rm ad}$ the adiabatic gradient. 
The equations are solved, using the pressure as an independent variable, for each layer separately, and altogether provide the internal structure of the planet. 

The temperature profile in the different layers follows an adiabat, with some temperature discontinuity at the transitions between the layers. 
At every transition, the temperature discontinuity follows the values suggested by Grasset et al. (2010), namely the 
temperature increases by 300 K at the inner/outer mantle transition, and by 1200 K at the mantle/water layer interface. The temperature is assumed to be 
continuous between the planetary surface and the gas envelope. As already demonstrated in different publications (e.g., Sotin et al. 2007, 
Seager et al. 2007, Grasset et al. 2010, Valencia et al. 2010), the planetary radius, at a given mass, hardly depends on the thermal profile. 
Given the relatively low depth of the water layer, we assume that this latter is isothermal. Finally, the gas envelope is treated as in Pierrehumbert and Gaidos (2011),
and similar to the one developed in Viktorowicz and Ingersoll (2007). For a given choice of the surface conditions, we follow an adiabat
until the skin temperature ($T_{\rm skin} = T_{\rm surf} /  2^{1/4}$, where $T_{\rm surf}$ is the surface temperature) is reached. 
Finally, we have also considered more detailed models of the planetary envelope including, for some of them, the irradiation from the star in a two-stream approach (see Guillot 2010).

The boundary conditions are as follows: we specify a surface temperature and pressure, as well as a central pressure. In addition, we assume that the pressure at the bottom of the ocean is the
equilibrium pressure, at the ice VII/water ice transition. This equilibrium pressure is given by a Simon law, $P = P_0 + \alpha_S \left( ( {T \over T_0} )^{\beta_S} -1 \right)$, with parameters given 
in Table \ref{table_Simon_water}. The pressure at the bottom of the ocean is computed based on the choice of the surface temperature and pressure. The thermal profile in the outer mantle is then 
computed, following an adiabat, until the transition to the inner mantle, at a pressure on the order of 22 GPa. Again, this transition depends on the temperature, following a similar law as for 
the ice/water transition: $P=P_0 + \alpha_S \left( T - T_0 \right)$, with parameters given in Table \ref{table_Simon_perov}. Finally, the temperature profile in the inner mantle and the core is
 computed following an adiabat until the assumed central pressure is reached. Once the thermal profile has been constructed, we integrate the internal structure equations, 
 using the pressure as an independent variable. The only unknown in the model is therefore pressure at the core/inner mantle interface, if there is enough Fe to 
 allow the presence of an iron core. We therefore use an iterative method to find the transition pressure that allows us to match the assumed composition of the planet (Fe/Si ratio, in particular).
 Once the internal structure of the non gaseous part of the planet has been derived, we compute  the structure of the atmosphere by following an adiabat starting from the 
 surface temperature and pressure, until the skin temperature is reached, following the method of Pierrehumbert and Gaidos (2011). The structure of the atmosphere is then used to derive the chord 
 radius, corresponding to the place where the chord optical depth (the one that is observed by transit) is equal to 1. The opacity used for this computation is given by Bell and Lin (1994) as a  function of the pressure 
 and temperature in the gas phase. Note that the Bell and Lin (1994) opacity is mainly due to grains, and probably represents an upper limit of the real opacity (see e.g., Mordasini et al. 2012b and references therein). Again, this approach will determine an upper boundary of the radius. To test the effect of opacity, we have computed some additional models, using the opacity derived by Freedman et al. (2008),
 with the low metallicity table.

\begin{table}
\caption{Parameters of the ice VII/liquid water transition}
\begin{center}
\begin{tabular}{cccc}
\hline\noalign{\smallskip}
 $P_0$ (GPa) & $T_0$ (K) & $\alpha_S$ (GPa) & $\beta_S$ \\
\noalign{\smallskip}
\hline\noalign{\smallskip}
 2.216 & 355 & 0.534  & 5.22 \\
\hline
\end{tabular}
\end{center}
\label{table_Simon_water}
\end{table}

\begin{table}
\caption{Parameters of the perovskite/olivine transition}
\begin{center}
\begin{tabular}{ccc}
\hline\noalign{\smallskip}
 $P_0$ (GPa) & $T_0$ (K) & $\alpha_S$ (GPa/K) \\
\noalign{\smallskip}
\hline\noalign{\smallskip}
25 & 800 & -0.0017  \\
\hline
\end{tabular}
\end{center}
\label{table_Simon_perov}
\end{table}

By following this procedure, \textit{i.e.} varying the central pressure, we obtain a set of internal structure models in which each value of the central pressure provides a different total mass. 
To obtain a mass-radius relationship, we vary the central pressure. Finally, since we are interested in the radius of potentially-habitable planets, we assume that the surface temperature 
varies in the range from 275K to 375K, and the surface pressure  varies in the range of $10^4$ to $10^9$ Pa. This range of surface conditions is rather arbitrary (and rather extended), 
and has been chosen to encompass what is believed to represent the range of surface conditions under which life could exist. The present study, however, could be easily extended 
to other surface conditions.

The derivation of the internal structure requires the specification of the equation of state (EOS), as well as the adiabatic gradient. The equations giving the pressure as a function of temperature and 
density are similar to the equations used by Sotin et al. (2007) for the inner and outer mantle, and are reproduced here for the sake of completion. We refer the reader to Sotin et al. (2007) for more 
details and justification of the use  of these EOS. For the inner mantle, the EOS is given by the Mie-Gruneisen-Debye formulation (see Poirier 2000):
\begin{equation}
P = P(\rho,T_0) + \Delta P
\end{equation}
\begin{equation}
\Delta P = \gamma \rho \left( E(T)-E(T_0) \right)
\end{equation}

\begin{eqnarray}
\begin{array}{l}
 P(\rho,T_0) = {3 \over 2} K_0 \left[ \left( {\rho \over \rho_0} \right)^{7/3} -  \left( {\rho \over \rho_0} \right)^{5/3} \right] \\
\qquad  \qquad \qquad \qquad \times \left( 1 - {3 \over 4} \left( 4-{K_0}^\prime \right) \left[ \left( {\rho \over \rho_0 }\right)^{2/3} -1 \right] \right)
\end{array}
\end{eqnarray}
and
\begin{equation}
E = { 9 n \over \mmol} P \left( { T \over \tetad  } \right)^3 \int_0^{\tetad / T} {x^3 e^x \over \left( e^x - 1 \right) } dx
\end{equation}
where $n$ is the number of atoms in the considered compound.
The Debye temperature $\tetad$ is given by 
\begin{equation}
\tetad = \theta_{\rm D,0} \left( \rho \over \rho_0 \right)^\gamma
\end{equation}
and $\gamma$ is given by $\gamma = \gamma_0 \left( \rho \over \rho_0 \right)^{-q}$. 

For the outer mantle and the liquid water layer, the EOS is given by the Birch-Murnaghan of third order formulation:
\begin{eqnarray}
\begin{array}{l}
P = {3 \over 2} K_{T,0}^0 \left[ \left( {\rho \over \rho_{T,0} } \right)^{7/3} -  \left( {\rho \over \rho_{T,0}} \right)^{5/3} \right] \\
\qquad  \qquad \qquad \qquad  \times \left(1 - {3 \over 4} \left( 4-{K_0}^\prime \right) \left[ \left( {\rho \over \rho_{T,0} }  \right)^{2/3} -1 \right] \right)
\end{array}
\end{eqnarray}
where $K_{T,0}^0 = K_0 + a_P (T -T_0)$, $\rho_{T,0}$ is given by
\begin{equation}
\rho_{T,0} = \rho_0 \exp \left( \int_{T0}^T \alpha (x,0) dx \right)
\end{equation}
and $\alpha(T,0) = a_T + b_T T - c_T / T^2$.
In these formulas,  $P_0$, $T_0$, and $\rho_0$ are the reference pressure (1 bar), temperature (300K) and corresponding density, and the values of the other parameters are given
in Tables \ref{table_MGD} and \ref{table_BM}. Note that since we use ${K_0}^\prime = 4$ for water,  the EOS is of second order only in this case.

\begin{center}
\begin{table*}[ht]
\caption{Parameters of the Mie-Gruneisen-Debye equation}
\begin{center}
\begin{tabular}{lccccccccc}
\hline\noalign{\smallskip}
 Specie & $\rho_0$ (g/cm$^3$) & $T_0$ (K) & $K_0$ (GPa) & $K_0^\prime$ (GPa/K) & $\theta_{D,0}$ (K) & $\gamma$ & $q$ & $\mmol$ (g) & $n$ \\
\noalign{\smallskip}
\hline\noalign{\smallskip}
MgO & 3.584 & 300 & 157 & 4.4 & 430 & 1.45 & 3 & 40.3 & 2 \\
MgSiO$_3$ & 4.108 & 300 & 263 & 3.9 & 1017 & 1.96 & 2.5 & 100.4 & 5\\
FeO & 5.864 & 300 & 157 & 4.4 & 430 & 1.45 & 3 & 80.1 & 2 \\
FeSiO$_3$ & 5.178 & 300 & 263 & 3.9 & 1017 & 1.96 & 2.5 & 131.9 & 5 \\
Fe & 8.334 & 300 & 174 & 5.3 & - & 2.434 & 0.489 & 55.8 & - \\
FeS & 4.9 & 300 & 135 & 6 & 998 & 1.36 & 0.91 & 87.9 & 2 \\
\hline
\end{tabular}
\end{center}
\label{table_MGD}
\end{table*}
\end{center}

\begin{center}
\begin{table*}[ht]
\caption{Parameters of the Birch-Murnaghan equation}
\begin{center}
\begin{tabular}{lcccccccc}
\hline\noalign{\smallskip}
 Specie & $\rho_0$ (g/cm$^3$) & $T_0$ (K) & $K_0$ (GPa) & $K_0^\prime$ (GPa/K) & $a_T$ (K$^{-1}$) & $b_T$ (K$^{-2}$) & $c_T$ (K) & $a_P$ (GPa K$^{-1}$)  \\
\noalign{\smallskip}
\hline\noalign{\smallskip}
liquid water & 1 & 300 & 2.2 & 4 & 0 & 0  & 0 & 0 \\
Mg$_2$SiO$_4$ & 3.222 & 300 & 128 & 4.3 & 2.832e-5 & 7.58e-9 & 0 & -0.016 \\
Mg$_2$Si$_2$O$_6$ & 3.215 & 300 & 111 & 7 & 2.86e-5 & 7.2e-9 & 0 & 0 \\
Fe$_2$SiO$_4$ & 4.404 & 300 & 128 & 4.3 & 2.832e-5 & 7.58e-9 & 0 & -0.016 \\
Fe$_2$Si$_2$O$_6$ & 4.014 & 300 & 111 & 7 & 2.86e-5 & 7.2e-9 & 0 & 0 \\
\hline
\end{tabular}
\end{center}
\label{table_BM}
\end{table*}
\end{center}

Finally, we use the EOS derived by Belonoshko (2010) for pure Fe, which is similar to the Mie-Gruneisen-Debye EOS, but has a different thermal pressure term:
\begin{eqnarray}
\begin{array}{l}
P = {3 \over 2} K_{T,0}^0 \left[ \left( {\rho \over \rho_{0} } \right)^{7/3} -  \left( {\rho \over \rho_{0}} \right)^{5/3} \right] \\
\qquad  \qquad \qquad \qquad  \times \left(1 - {3 \over 4} \left( 4-{K_0}^\prime \right) \left[ \left( {\rho \over \rho_{0} }  \right)^{2/3} -1 \right] \right) \\
\qquad \qquad \qquad \qquad + 3 R \gamma (T-T_0) \times M/ \rho
\end{array}
\end{eqnarray}
where the parameters are given in Table \ref{table_MGD}, and $\gamma$ has the same definition as for the Mie-Gruneisen-Debye EOS. This EOS for Fe
has been preferred to the one used by Sotin et al. (2007), as it allows to reproduce  high-precision volumetric  experiments of iron under high pressure 
and temperature closely (see Belonoshko 2010).

The adiabatic gradient is computed as 
\begin{equation}
{d T \over d P} ={  \gamma T \over K_S } 
\end{equation}
where $K_S$ is the adiabatic bulk modulus, which is related to the isothermal bulk modulus $K_T$ through
$K_S = \left( 1 + \gamma \alpha T \right) K_T$, $\alpha$ being the thermal expansion coefficient. In the case of the inner and outer mantle, we use the values derived by Katsura et al. (2010):
\begin{equation}
\alpha = \left( \alpha_0 + (T-T_0) a_1\right) \left( {\rho_0 \over \rho } \right)^{\delta_T}. 
\end{equation}
In these formulas, $a_1$ is the derivative of the  expansion coefficient with respect to the temperature, and $\delta_T$ provides the dependance of $\alpha$ with respect to the density. 
The parameter $\tetad$ is the Debye temperature, which depends on the density (see above). The numerical values of the different parameters are given in Table \ref{table_adiabat}.
In the case of the iron core, the value of the thermal expansion coefficient are directly computed from  Belonoshko's EOS (Belonoshko 2010).  Finally, we assume the water layer is isothermal.
This assumption has no practical consequences given the low depth of the ocean (see Fig. \ref{ocean}).

\begin{center}
\begin{table*}[ht]
\caption{Parameters used to compute the adiabatic gradient}
\begin{center}
\begin{tabular}{lccccccccc}
\hline\noalign{\smallskip}
 layer & $\delta_T$ & $a_1$ & $\alpha_0$ & $\gamma_0$ & $q$ & $\theta_0$ (K) & $\mmol$ (g) & $T_0$ (K) & $\rho_0$ (g/cm$^3$) \\
 \hline\noalign{\smallskip}
outer mantle & 7.2 & 1.6e-8 & 2.56e-5 & 1.26 & 2.9 & 760. & 140.7 & 300 & 3.222 \\
inner mantle & 5.8 & 1e-8 & 2.61e-5 & 1.96 & 2.5 & 1017 & 40.3 & 300 & 4.108 \\
\noalign{\smallskip}
\hline
\end{tabular}
\end{center}
\label{table_adiabat}
\end{table*}
\end{center}

As already mentioned above, the planetary composition is set by three parameters: the Mg/Si, Fe/Si molar ratios, and the composition of the planetary atmosphere. 
The water content is not an input parameter, but is derived from the constraint that the pressure at the bottom of the ocean is the crystallization pressure. These quantities are assumed, 
with one exception presented in Sect. \ref{max_all}, to be homogeneous in the entire planetary mantle.  

\section{Results}
\subsection{Test of the code: internal structure of an Earth-like planet}
\label{1mearth}

\begin{figure}
  \center
  \includegraphics[height=0.3\textheight]{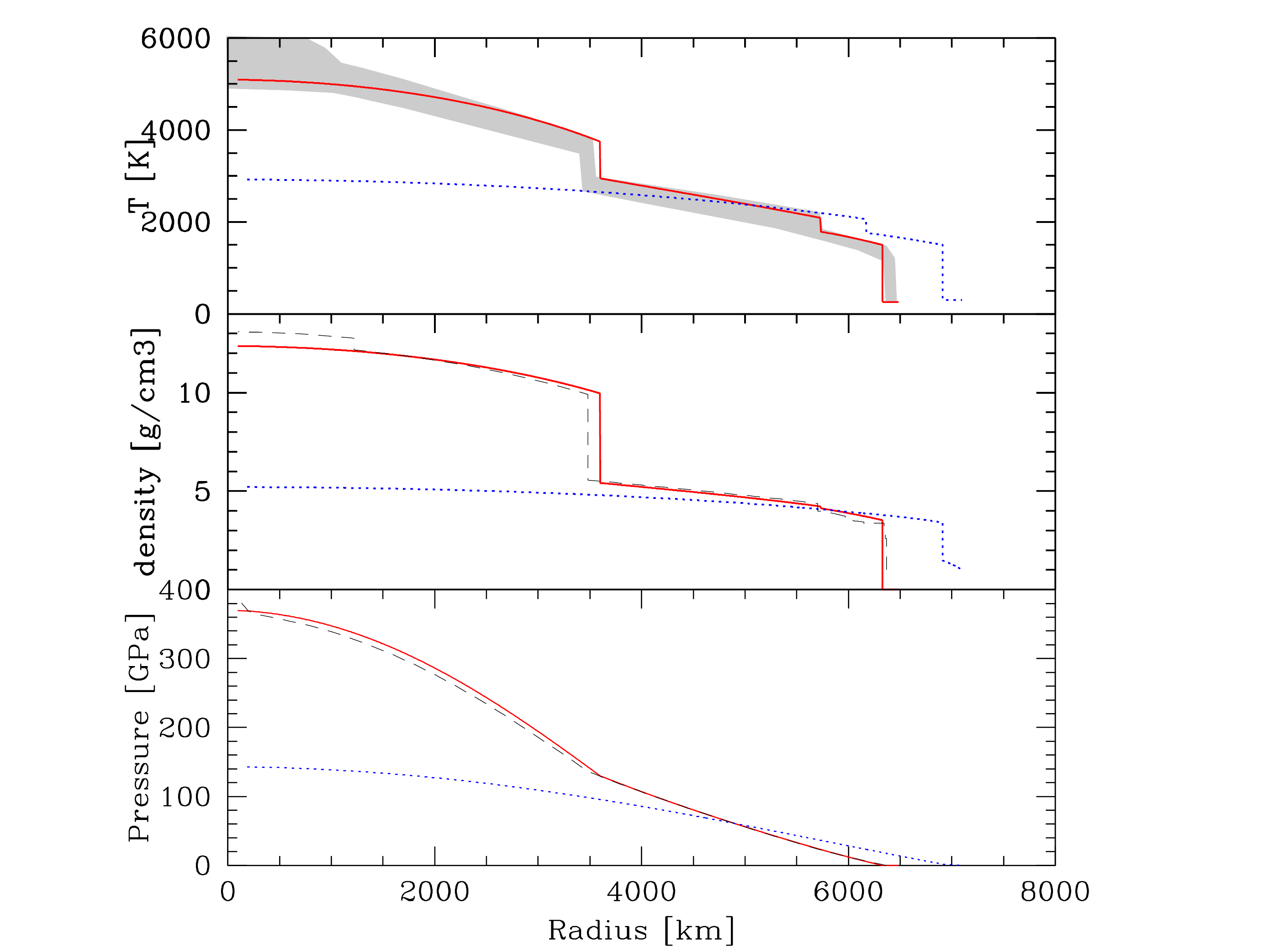}
  \caption{Temperature, density, and pressure (top, middle, bottom panels) as a function of radius for two models. Solid red lines: $1 \mearth$ planet with $T_{\rm surf}$ = 300 K, $P_{\rm surf} = 1$ bar, without ocean, and quasi-solar composition (see text).   Dotted blue lines: the same  planet, but corresponding to the overall maximum models (iron free,  inner mantle made of MgO, outer mantle made of Mg$_2$Si$_2$O$_6$, with maximum  ocean depth). In the middle and bottom panels, the black dashed lines indicate the Preliminary Reference Earth Model of Dziewonski \& Anderson (1981). Note that no temperature is given in this model. Instead, the grey area reproduces the 
  likely temperature of Earth as mentioned in Sotin et al. 2007.}
  \label{internal_structure}
\end{figure}

As an example and test of our code, in Fig. \ref{internal_structure} we present the temperature, pressure, and mass as a function of the radius, for an Earth-like planet. The surface conditions
are T=300K and P = 1 bar. The composition of the planet is similar to model 1 in Sotin et al. (2007), namely Mg/Si = 1.131 and Fe/Si = 0.986. We  assume, for the sake of 
comparison, that 13\% of the core is made of FeS (treated using the Mie-Gruneisen-Debye EOS), and 87\% is made of pure Fe. We refer the reader to Sotin et al. (2007) for the justification of this 
slightly non-solar composition. To have a planet of 1 $\mearth$, the central pressure is equal to 370 GPa, the pressure at the core-mantle transition is equal to 130 GPa, and the pressure at 
the inner/outer mantle transition is equal to 23.31 GPa. The corresponding temperatures are  5092K, 3747K and 1789 K respectively. The mass obtained using these parameters is 1.007 $\mearth$, 
and the radius of the solid part is  $0.993 \rearth$. The pressure, density, and temperature as a function of the radius are   similar to the Preliminary Reference Earth Model
 (Dziewonski \& Anderson, 1981), except for the inner/outer core  structure that is not modeled in our code. This difference in the core structure (and mean density) also explains the slight shift (in
 radius) of the core-mantle boundary. For this type of planet, the maximum depth of the ocean (that is not taken into
  account in the numbers shown above) would be 150km, corresponding to a mass of 0.016 $\mearth$, which is more than 100 times the inventory of water on Earth. 

In the same figure, we also present the case corresponding to the maximum radius, an iron free planet, whose inner core i made of MgO, and whose outer core is made of Mg$_2$Si$_2$O$_6$.
In this second example,  the inner and outer mantle composition are not the same and the overall Mg/Si ratio is equal to 2.86, nearly three times the solar value.  For this planet, the mass of the ocean is 
0.0243 $\mearth$ and its depth is 188 km. The ocean is larger because the gravity is weaker (the planet is less dense than in the previous case). As a consequence, the ice VII/liquid water transition 
is reached at higher depth. The pressure at the bottom of the ocean and the pressure at the inner/outer mantle transition are the same, since they only depend on the surface temperature, which is the same 
as in the previous case. The central pressure and temperatures are  143 GPa and 3722 K respectively. Finally, the mass of the gas envelope is $1.45 \times 10^{-7} \mearth$, the transit radius of the 
planet is 1.30$\rearth$, and the radius of the solid/liquid planet (called inner planet in the following) is 1.11 $\rearth$.  

\subsection{Overall maximum radius of planets}
\label{max_all}

To derive the overall maximum radius of a potentially habitable planet, in the sense we defined in the introduction, we first assume, unrealistically, that the planets do not contain any iron. 
Moreover, we only consider the less dense phases in the inner and outer mantle, namely MgO (inner core) and Mg$_2$Si$_2$O$_6$ (outer core). It should be noted that in this case, 
the Mg/Si ratio is different in the inner and outer core. We admit that this kind of planetary structure is likely to be unrealistic, but we only considered it to provide the overal maximum radius
of a planet at a given mass.

\begin{figure}
  \center

  \includegraphics[height=0.25\textheight]{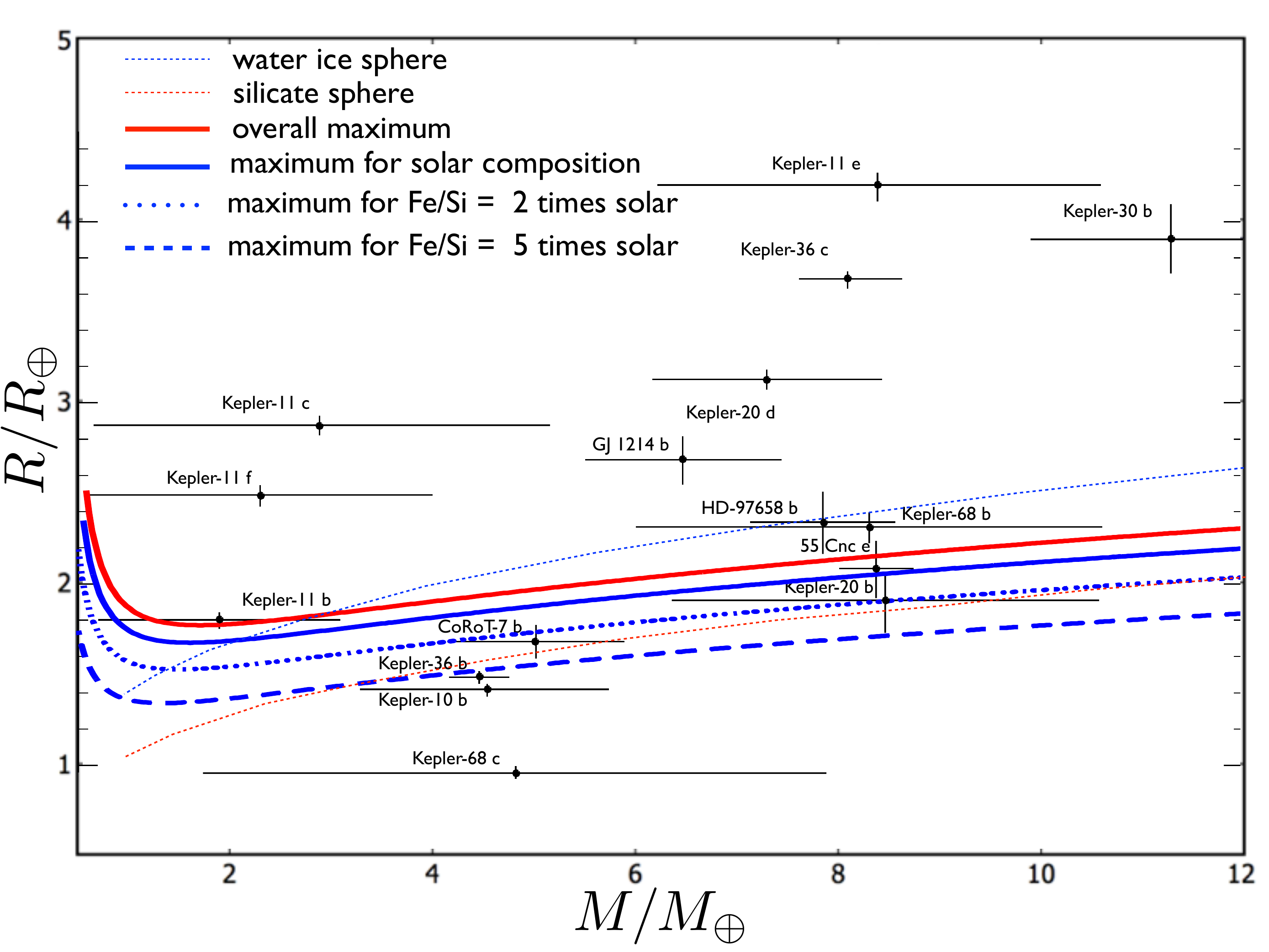}
    
  \caption{Mass \textit{versus} maximum radius relationship for different composition of the planetary interior. The heavy red solid line corresponds to an iron-free planet, with an inner mantle made of MgO, 
  and an outer mantle made of Mg$_2$Si$_2$O$_6$. The blue lines are computed assuming a solar Mg/Si and a Fe/Si equal to the solar value (solid line), two times the solar value (dotted line), and five
  times the solar value (dashed line). The mass-radius relationships for a sphere of silicates and a sphere of water are indicated by thin solid lines (red and blue respectively), and are taken from Wagner et al.
  (2011). The parameters of some transiting planets are taken from exoplanets.org the 2013 September 16.}
  \label{MR}
\end{figure}

Using the model we have described above, we now derive the maximum radius a planet can have to harbor both a C-cycle and surface conditions in
the range defined above. The overall maximum is obtained, as mentioned
above, assuming that the planet has a very peculiar composition, where only low density minerals - MgO for the inner mantle and Entstatite (Mg)
for the outer mantle - are present in the planetary interior. Figure \ref{MR} shows the resulting mass-radius relation we obtain, with some transiting planets also plotted in the figure (Data were taken from exoplanet.org). 
As can be see from the figure, the maximum radius increases from 1.8 $\rearth$ to 2.3 $\rearth$ for planets ranging from 1 $\mearth$ to 10 $\mearth$.
Lower mass planets can have a larger radius, the increase of the mass-radius relationship being the result of the reduced gravity.
Various studies (Mordasini et al. 2012a and references therein) have described such behavior of the mass-radius relation. 
As can be seen in Fig. \ref{MR}, a large fraction of the planets represented
in the figure are not habitable, based on the criteria we have described in the introduction (it is anyway obvious that all these planets are not habitable due to their proximity 
to the central star). 

\subsection{Maximum radius for different compositions}
\label{max_fe}

Assuming some values of the Fe/Si and Mg/Si ratios in the planetary interior (these values reflect approximate ratios in the central star, see Thiabaud et al. and Marboeuf et al. , in prep), we can repeat the same
calculations. In this calculation, the fraction of the different compounds (Fe, perovskite, wustite, olivine, entsatite) is computed assuming that the mantle is homogeneous: the Mg/Si and Fe/Si ratios are the same 
in the inner and outer mantle. Since we are interested in the lowest  density planet, the fraction of iron in the mantle is maximized. 
It is only necessary to assume the existence of a central iron core for Fe/Si ratios larger than 1.13 (in the case of a solar Mg/Si ratio).

\begin{figure}
  \center

  \includegraphics[height=0.3\textheight]{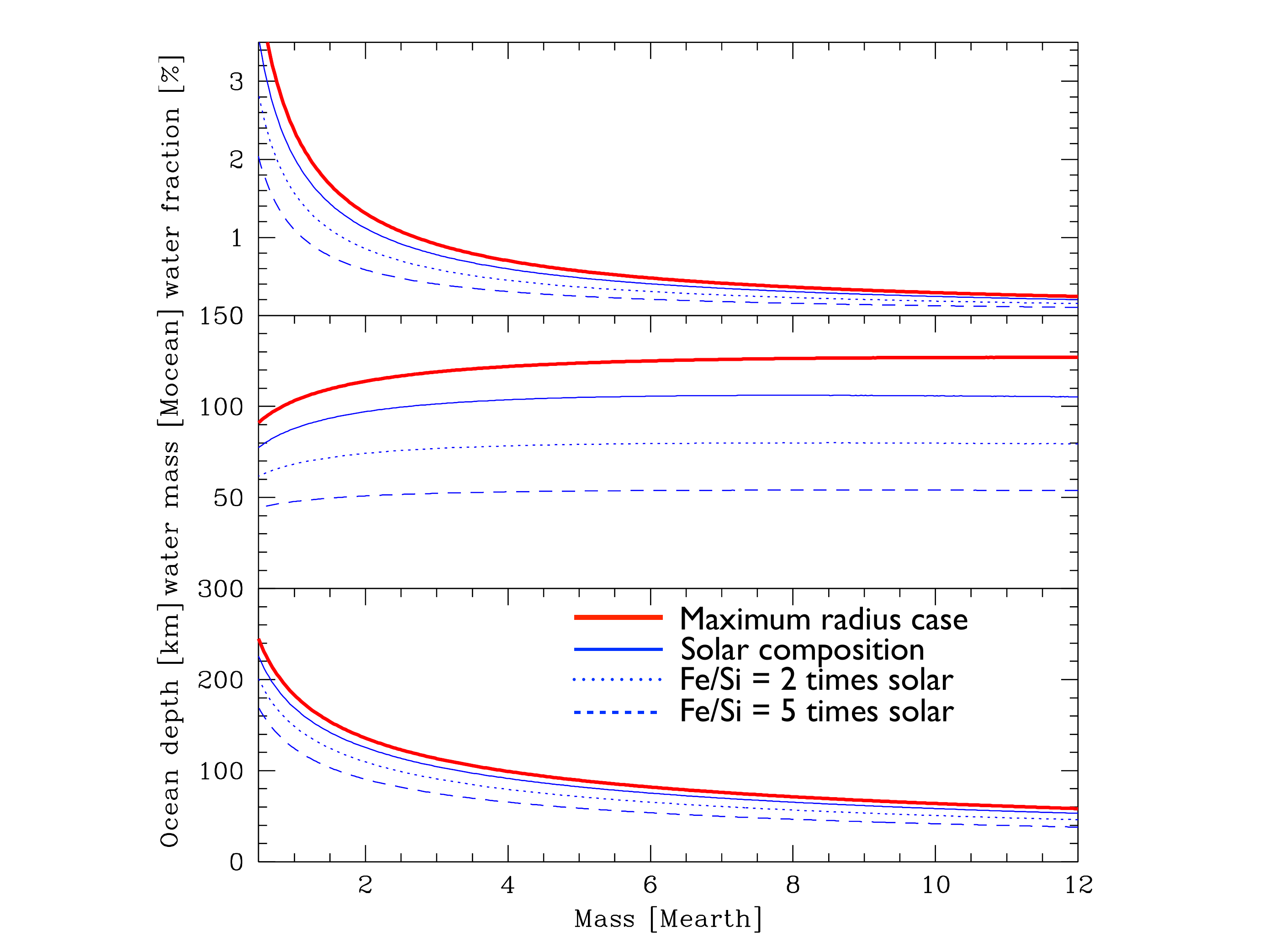}
    
  \caption{Characteristic of the maximum ocean as a function of the planetary mass, for different models. The top, middle and bottom panels present respectively the water fraction (considering only the ocean),
  the ocean mass relative to the ocean mass on Earth ($\mocean = $ 2.3 $10^{-4}\mearth$), and the ocean depth. The heavy red solid line is in the case of an iron free planet, with an inner mantle made of MgO, 
  outer mantle made of Mg$_2$Si$_2$O$_6$. The blue lines are computed assuming a solar Mg/Si, and a Fe/Si equal to the solar value (solid line), two times the solar value (dotted line), and five
  times the solar value (dashed line).}
  \label{ocean}
\end{figure}

Fig. \ref{MR} shows the resulting mass-radius relation for different values of the Fe/Si ratio, namely 1, 2, and 5 times the solar value (0.8511). The Mg/Si ratio is kept at its solar value (1.0243)
for all the simulations. As can be seen, the planets with solar composition have a maximum radius less than 0.1 $\rearth$ smaller than the overall maximum radius derived in the previous
section. The shift in radius is almost independent of the planetary mass, but depends on the Fe/Si ratio: planets are around 0.2 $\rearth$ smaller (than the overall maximum) for 
Fe/Si = 2 Fe/Si$)_\odot$, and 0.4 $\rearth$ smaller for Fe/Si = 5 Fe/Si$)_\odot$. Correspondingly, the maximum mass of water decreases from $\sim 125 \mocean$ to 105, 80 and 55 $\mocean$
for Fe/Si ratios equal to 1, 2, and 5 times the solar value respectively (see Fig. \ref{ocean}).

It is also interesting to consider the maximum mass of the ocean on the planets. Fig. \ref{ocean} shows the maximum ocean mass, and the maximum water fraction as a function
of the planetary mass. Note that the maximum water fraction is based on the ocean mass only and does not include water that in reality could be  incorporated in the planetary
interior. The maximum water mass is found to be nearly independent of the planetary mass, except for low mass planets. Accordingly, the maximum water fraction decreases with planetary 
mass, from 2 \% for an Earth mass planet to 0.2 \% for a 12 $\mearth$ planet.

\subsection{Effect of the atmosphere model}
\label{atm_model}

\begin{figure}
  \center

  \includegraphics[height=0.3\textheight]{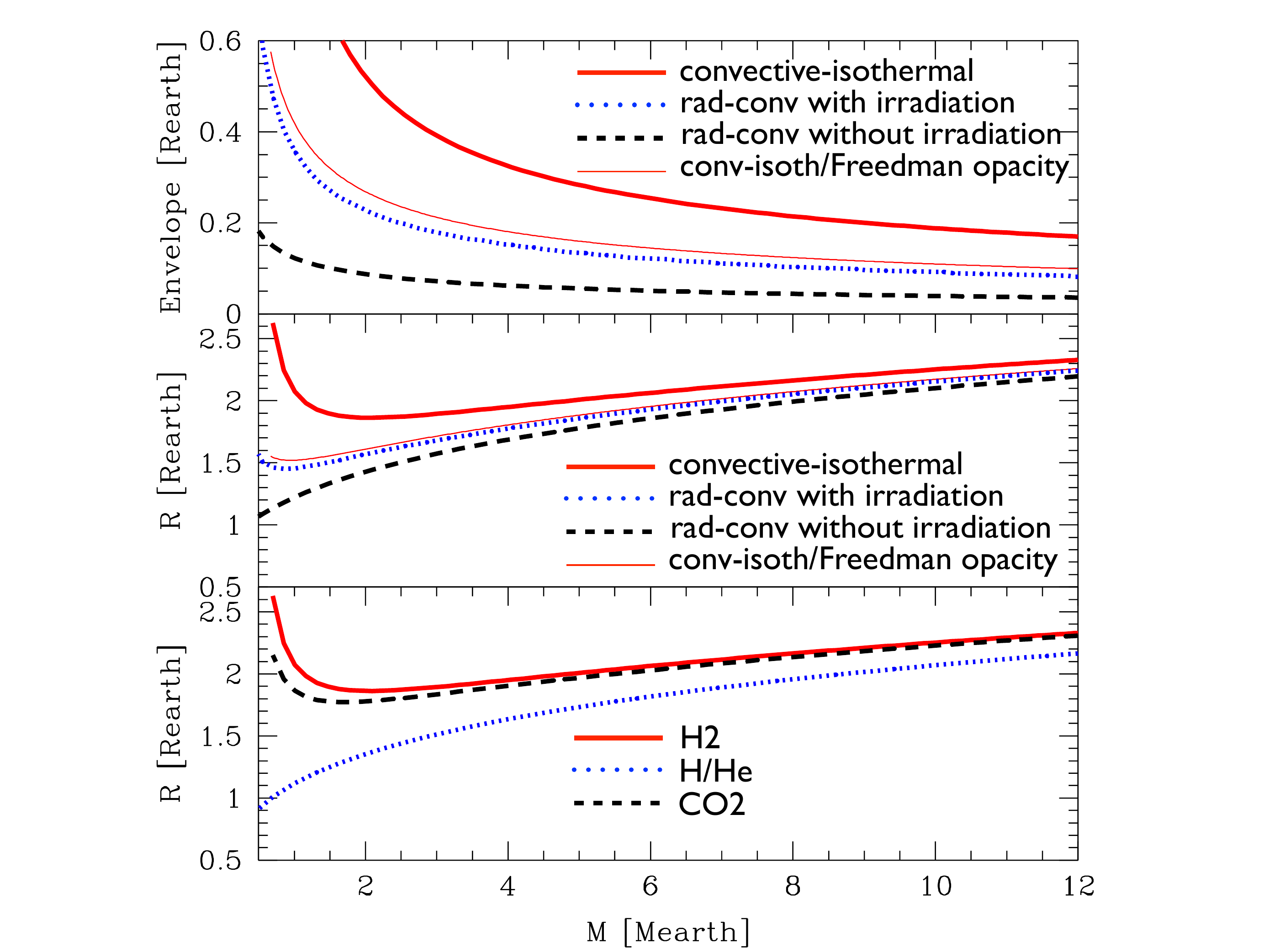}
    
  \caption{{\bf Top}: Depth of the gas envelope (in Earth radius), as a function of the planetary mass, for the three models presented in Sect. \ref{atm_model}, for a pure H$_2$ envelope. The red solid line is the convective-isothermal model, the black dashed line is the radiative-convective model without  irradiation, and the blue dotted line is the radiative-convective model with irradiation effect. The red thin solid line
  is similar to the red solid line, except that we use the opacities of Freedman et al. (2008), for the low metallicity case, to derive the planetary radius. {\bf Middle}: The mass-radius relationship for the four same models. {\bf Bottom}: The mass-radius relationship
  for the  convective-isothermal model, assuming three compositions. The red solid line is for pure H$_2$, the black dashed line is for a mixture of H and He in solar proportions, and the blue dotted line is for CO$_2$ (not including condensation). For 
  the three panels, the planetary interior model is computed as in Sect. \ref{max_all}, and the surface temperature and pressure are equal  to 375K and 1 GPa respectively.}
  \label{atmosphere}
\end{figure}

An important contribution to the planetary radius comes from the gas envelope. The top panel of Fig. \ref{atmosphere} shows the envelope depth for different models, and for the maximum radius, iron-free planet, similar to the case discussed in Sect. \ref{max_all}. We have considered three different models in this figure. In the first one (named later convective-isothermal), as we explained before, the pressure-temperature
profile is assumed to follow an adiabat, starting with a given surface temperature and pressure, where the temperature is limited to the skin temperature (see Pierrehumbert and Gaidos 2011).

In the second model, we proceed as follows: for a given mass and radius of the inner planet (excluding any gas envelope), we have computed a set of envelope structure models, assuming different values
of the total (solid + gas) planetary radius, and a planetary luminosity. The standard internal structure equations (see e.g., Alibert et al. 2005, Alibert et al. 2013) are then solved from outside in, namely from the 
assumed total radius, to the radius of the inner planet (solids+ocean). The luminosity is assumed to be uniform in the gas envelope and is a free parameter, which is given by an equivalent internal temperature, as presented in Guillot 2010. The outermost pressure is assumed to be a small arbitrary value ($10^{-4}$ Pa), and the outermost temperature is equal to the standard outer temperature in the Eddington 
approximation ($T_{\rm out} = T_{\rm int} / 2^{1/4}$, where $T_{\rm int}$ parametrizes the planetary luminosity). In the ensemble of models we compute, we finally select the subset of planetary
radius and planetary luminosity (or $T_{\rm int}$) that allow us to match some values of the surface pressure and temperature. We computed 90000 models for an envelope depth spanning values
from one scale height to 200 scale height (the scale height being computed for the planetary outermost temperature), and for $T_{\rm int}$ spanning values from 1K to 300 K.
The maximum radius we obtain with this procedure is  the
maximum radius a planet can have for a given set of surface temperature and pressure (375K and 1 GPa for the models considered in Fig. \ref{atmosphere}), 
as well as a given radius and mass of the solid planet. It appears that the mass of the envelope is always negligible compared to the total mass of the planet, at least for surface pressure 
and temperature compatible with the presence of liquid water at the planetary surface, and one can to a very high accuracy identify the mass of the inner planet to the total planetary mass. 

Finally, the third model is computed in a way similar to the second one, but taking the irradiation of the parent star into account, following the two-stream model of Guillot et al. (2010). The irradiation
temperature is assumed to be equal to the one at the present Earth location, namely $T_{\rm irr} =  255$K. In this case, we again compute a set of envelope structure models, and select the ensemble
of radii and planetary luminosity that allow us to match any given surface pressure and temperature. The opacity in the visible and IR range are assumed to be equal to the values quoted in Guillot
et al. (2010, see caption of Fig. 5), namely $\kappa_{\rm vis} = 2 \times 10^{-3} {\rm cm}^2$/g and $\kappa_{\rm IR} = 10^{-2}  {\rm cm}^2$/g. The factor $f$ is taken 
equal to 0.25, meaning that we assume a redistribution of the incoming energy over the whole planetary surface, which is consistent with our 1D spherically-symmetric models. 
 Finally note that the second model is indeed a particular case of the third one, with an irradiation temperature equal to 0, and with non-constant opacities (Bell and Lin 1994).
                 
We present in the middle panel of Fig. \ref{atmosphere} the different mass-radius relationships for the three models, in the case described in Sect. \ref{max_all}, and for an envelope gas made of H$_2$. As shown on the figure, the difference in planetary radius is  small when considering the three models, except for low mass planets, where the low gravity enhances the effect of gas composition and temperature gradient.
This similarity comes from the fact that the temperature change between the outermost layers and the surface is rather small. As a consequence, the exact temperature profile (which is the main difference between the  three envelope models) has a small effect.
In addition, we have computed the same models,  using the opacities of Freedman et al. (2008), as the Bell and Lin (1994) opacities represent likely an over estimation of the actual opacity. As expected, this
translates to a smaller envelope depth for the same model (see the upper and middle panels of Fig. \ref{atmosphere}).

For the first (convective/isothermal) model, we have computed the maximum radius considering a different atmospheric composition, namely pure H$_2$, a mixture
of H and He in solar proportions, and CO$_2$ (molecular weight of 44  g/cm$^3$, and $\gamma = 1.29$). Note that in the case of CO$_2$, we do not take any possible condensation into account,
and the profile is therefore very academic. Taking more realistic atmospheric profiles into account will be the subject of future work. Nevertheless, these different examples are used to show the effect
on the planetary maximum radius and are believed to  bracket the reality. 

As can be seen on lower panel of Fig. \ref{atmosphere},  increasing the molecular weight decreases the planetary radius, as expected. By measuring the slope of the
planetary transit in two wavelengths in the visible, one can deduce the mean molecular weight in the planetary atmosphere (at least the upper parts, see Benneke and Seager 2012). By assuming that this value represents the molecular weight in the entire envelope, one could therefore derive a more stringent value of the planetary maximum radius. 

\section{Discussion and conclusion}
\label{conclusion}

We have derived the maximum radius of  planets in the Earth to Super-Earth regime under the hypothesis that both the surface conditions at the planetary surface
lie in the liquid domain of the water phase diagram, and that the pressure at the bottom of a putative global ocean is lower than the liquid water/ice VII transition (approximately
2.4 GPa). As we have argued in the introduction, if these two conditions are indeed necessary for habitability, the maximum radius we derive delimits a region above
which (in terms of radius) planets are very likely not habitable.

For this calculation, we have constructed internal structure models for the planet, made of a central iron core, a silicate mantle (itself divided in two layers), a water layer, and a gas envelope.
Our model for the inner planet follows the model originally developed by Sotin et al. (2007), and further improved by Grasset et al. (2010), with some differences related to the
assumed EOS iron, where we have used the most recent EOS developed by Belonoshko (2010). For the gas envelope, we have considered three models, one convective/isothermal
model, one radiative/convective model, and a two-stream irradiated model following the model of Guillot (2010). We have checked that our model reproduces the Preliminary 
Reference Earth Model  (Dziewonski \& Anderson, 1981), except for the inner/outer core  structure, which is not included in our code. 

We have  first considered the case of planets devoid of Fe surrounded by a pure H$_2$ envelope first. Under these (rather unrealistic) conditions, planets larger than 1.8 to 2.3 $\rearth$,
for masses ranging from 2 to 12 $\mearth$, cannot harbor conditions compatible with both the presence of liquid water at the surface and a C-cycle. It is important to remember that this maximum radius is derived
under very extreme conditions (no iron, pure H$_2$ convective/isothermal envelope). Under more realistic conditions (e.g., a differentiated planet of super-solar composition, a gas envelope of higher 
molecular weight), the maximum radius for a given mass can be reduced by up to $\sim$ 0.5 $\rearth$. Correspondingly, the maximum fraction of water in planets must be smaller than $\sim 1 \%$ for planets
more massive than 1 $\mearth$. This maximum fraction decreases with the planetary mass as well as with the iron content of the planet.

To estimate the uncertainties in our model, we have varied the gas envelope model, and found that the resulting radius can vary by at most  $\sim 0.1 \rearth$ when considering
either the convective/isothermal model or the irradiated convective/radiative model. We have also checked that varying the IR and visible opacities in the irradiated models by factors up to 
2, or assuming no redistribution on the planetary surface (factor $f$ equal to 1) does not modify the resulting mass-radius relationships notably .
 In all the cases, the radius is found to be smaller than
in the simple model, and we are therefore confident that the maximum radius derived in Sect. \ref{max_all} represents a boundary above which the habitability of planets is very unlikely.
The thermal profile in planets is also unknown and has an effect on the resulting radius. Other studies (e.g., Seager et al. 2007, Grasset et al. 2010) have  shown however that the effect is of a few percent. We have indeed checked that a variation of a few percent of the adiabatic gradient
does not  change our results. 

 In our model we have considered that the ocean is made of pure water, which is an assumption that is probably not correct. Indeed, observations have shown that the Jupiter's icy satellites (whose structure could be seen as the one of a small ocean planet at large distance) contain a lot of volatiles, like CO$_2$ (see Hibbit et al. 2000, 2002) or NH$_3$ (see
Mousis et al. 2002). In addition, recent population synthesis models, based on our planet formation model (Alibert et al. 2005,
Mordasini et al. 2009a,b, Fortier et al. 2013, Alibert et al. 2013), show that planets that contain water also contain
similar volatiles (NH$_3$, CO$_2$, CO, etc...). The presence of such volatiles is known to modify the crystallization
process of water and could modify our results (see e.g., Spohn \& Schubert 2003). However, an appreciable change in the maximum radius we have 
derived would require that the crystallization pressure of water mixed with volatiles is substantially increased.
Future studies will be required to address  this issue properly. 

Finally, it is possible to derive a stronger constraint on the maximum radius of a planet, if one can determine both the abundance of key elements in the central star (in particular Fe, Si, and Mg), 
and the mean molecular weight of the gas envelope, using Rayleigh scattering observations (see Benneke and Seager 2012). Such a derivation  however relies on the assumption that the refractory composition of planets is similar 
to the one of the central star, an assumption that is supported by recent population synthesis models (Thiabaud et al, in prep). Using such models, transit observations such as the one that will be performed
by CHEOPS (see Broeg et al. 2013) or TESS (the Transiting Exoplanet Survey Satellite, see Ricker et al. 2010), complemented by ground-based or other space-based observations, it will be possible to select the best candidates for future habitability studies.

\acknowledgements

This work was supported by the European Research Council under grant 239605.

\end{document}